\documentclass[11pt]{article}
\usepackage{amsfonts,amssymb,a4,bm}
\usepackage{color}
\usepackage[all]{xy}
\usepackage{color}
\definecolor{Red}{cmyk}{0,1,1,0}
\def\theequation{\thesection.\arabic{equation}}

\newcommand{\qed}{\hfill\rule{3mm}{3mm}}
\newtheorem{teorema}{Theorem}
\newtheorem{defi}{Definition}
\newtheorem{lem}{Lemma}

\newtheorem{pro}{Proposition}

\makeatletter \@addtoreset{equation}{section} \makeatother
\begin{document}


\voffset=-1.5truecm\hsize=16.5truecm    \vsize=24.truecm
\baselineskip=14pt plus0.1pt minus0.1pt \parindent=12pt
\lineskip=4pt\lineskiplimit=0.1pt      \parskip=0.1pt plus1pt

\def\ds{\displaystyle}\def\st{\scriptstyle}\def\sst{\scriptscriptstyle}


\let\a=\alpha \let\b=\beta \let\ch=\chi \let\d=\delta \let\e=\varepsilon
\let\f=\varphi \let\g=\gamma \let\h=\eta    \let\k=\kappa \let\l=\lambda
\let\m=\mu \let\n=\nu \let\o=\omega    \let\p=\pi \let\ph=\varphi
\let\r=\rho \let\s=\sigma \let\t=\tau \let\th=\vartheta
\let\y=\upsilon \let\x=\xi \let\z=\zeta
\let\D=\Delta \let\F=\Phi \let\G=\Gamma \let\L=\Lmbda \let\Th=\Theta
\let\O=\Omega \let\P=\Pi \let\Ps=\Psi \let\Si=\Sigma \let\X=\Xi
\let\Y=\Upsilon\let\L\Lambda



\global\newcount\numsec\global\newcount\numfor
\gdef\profonditastruttura{\dp\strutbox}
\def\senondefinito#1{\expandafter\ifx\csname#1\endcsname\relax}
\def\SIA #1,#2,#3 {\senondefinito{#1#2}
\expandafter\xdef\csname #1#2\endcsname{#3} \else \write16{???? il
simbolo #2 e' gia' stato definito !!!!} \fi}
\def\etichetta(#1){(\veroparagrafo.\veraformula)
\SIA e,#1,(\veroparagrafo.\veraformula)
 \global\advance\numfor by 1
 \write16{ EQ \equ(#1) ha simbolo #1 }}
\def\etichettaa(#1){(A\veroparagrafo.\veraformula)
 \SIA e,#1,(A\veroparagrafo.\veraformula)
 \global\advance\numfor by 1\write16{ EQ \equ(#1) ha simbolo #1 }}
\def\BOZZA{\def\alato(##1){
 {\vtop to \profonditastruttura{\baselineskip
 \profonditastruttura\vss
 \rlap{\kern-\hsize\kern-1.2truecm{$\scriptstyle##1$}}}}}}
\def\alato(#1){}
\def\veroparagrafo{\number\numsec}\def\veraformula{\number\numfor}
\def\Eq(#1){\eqno{\etichetta(#1)\alato(#1)}}
\def\eq(#1){\etichetta(#1)\alato(#1)}
\def\Eqa(#1){\eqno{\etichettaa(#1)\alato(#1)}}
\def\eqa(#1){\etichettaa(#1)\alato(#1)}
\def\equ(#1){\senondefinito{e#1}$\clubsuit$#1\else\csname e#1\endcsname\fi}
\let\EQ=\Eq

\def\V{V}
\def\Rd{{\mathbb{R}^d}}

\def\dpr{\partial}


\def\\{\noindent}
\let\io=\infty

\def\VU{{\mathbb{V}}}
\def\ED{{\mathbb{E}}}
\def\GI{{\mathbb{G}}}
\def\Tt{{\mathbb{T}}}
\def\C{\mathbb{C}}
\def\LL{{\cal L}}
\def\RR{{\cal R}}
\def\SS{{\cal S}}
\def\NN{{\cal M}}
\def\MM{{\cal M}}
\def\HH{{\cal H}}
\def\GG{{\cal G}}
\def\PP{{\cal P}}
\def\AA{{\cal A}}
\def\BB{{\cal B}}
\def\FF{{\cal F}}
\def\TT{{\cal T}}
\def\v{\vskip.1cm}
\def\vv{\vskip.2cm}
\def\gt{{\tilde\g}}
\def\E{{\mathcal E} }
\def\I{{\rm I}}
\def\0{\emptyset}
\def\xx{{\V x}} \def\yy{{\bf y}} \def\kk{{\bf k}} \def\zz{{\bf z}}
\def\ba{\begin{array}}
\def\ea{\end{array}}  \def \eea {\end {eqnarray}}\def \bea {\begin {eqnarray}}
\def\xto#1{\xrightarrow{#1}}

\def\tende#1{\vtop{\ialign{##\crcr\rightarrowfill\crcr
              \noalign{\kern-1pt\nointerlineskip}
              \hskip3.pt${\scriptstyle #1}$\hskip3.pt\crcr}}}
\def\otto{{\kern-1.truept\leftarrow\kern-5.truept\to\kern-1.truept}}
\def\arm{{}}
\font\bigfnt=cmbx10 scaled\magstep1

\newcommand{\card}[1]{\left|#1\right|}
\newcommand{\und}[1]{\underline{#1}}
\def\1{\rlap{\mbox{\small\rm 1}}\kern.15em 1}
\def\ind#1{\1_{\{#1\}}}
\def\bydef{:=}
\def\defby{=:}
\def\buildd#1#2{\mathrel{\mathop{\kern 0pt#1}\limits_{#2}}}
\def\card#1{\left|#1\right|}
\def\proof{\noindent{\bf Proof. }}
\def\qed{ \square}
\def\reff#1{(\ref{#1})}
\def\eee{{\rm e}}
\def\be{\begin{equation}}
\def\ee{\end{equation}}

\title{Virial series for a system of classical particles interacting through a pair potential with negative minimum}

\author{\normalsize Aldo Procacci\\{\scriptsize Departamento de Matem{\'a}tica, Universidade Federal de Minas Gerais, Belo
Horizonte-MG, Brazil - aldo@mat.ufmg.br}
}

\maketitle

\begin{abstract}
In this note we revisit the recent developments concerning  rigorous results on the  virial series
of a continuous system of classical particles interacting via a stable and tempered pair potential and
we provide new lower bounds for its  convergence radius when the potential  has a strictly positive stability constant.
As an application we obtain a new estimate for the convergence radius of the virial series   of  the Lennard-Jones gas
which improves sensibly previous estimates present in the literature.
\vskip.3cm

%
%
\end{abstract}

\vskip.3cm
{\footnotesize

}

\numsec=1\numfor=1
\let\thefootnote\relax
\subsection*{1. Introduction: Model and results}
In this note we consider a system of classical identical  particles confined in a box $\L\subset \Rd$
interacting via a translational invariant and even pair potential $v: \Rd\to \mathbb{R}\cup\{{+\infty}\}$. This  system is described
in the Grand
Canonical Ensemble  by a probability measure on $\cup
_{n} \G_n(\L)$ where $\G_{n}(\L)~=~\{(x_1,\dots x_n)\in
\L^{dn}\}$ whose restriction to $\G_n(\L)$ is
$$
d\m(x_1,\dots,x_z)~=~ {1\over \Xi^v_\L (\b,z)}{z^{n}\over n!}
e^{ -\b \sum_{1\le i<j\le n}v(x_i-x_j)} dx_1\dots dx_n \Eq(2.1)
$$
where $\b~=~(kT)^{-1}$ is the inverse of the temperature in units of the  Boltzmann constant and $z$ is the (configurational) fugacity.  The normalization constant $\Xi^v_\L (\b,z)$, i.e.
 the grand canonical partition function is explicitly written as
$$
\Xi^v_\L (\b,z)~=~1+ \sum_{n=1}^{\infty}{z^n \over n!}
\int_{\L}dx_1\dots\int_{\L} dx_{n} ~e ^{-\b \sum_{1\le i<j\le n}v(x_i-x_j)} \Eq(2.2)
$$
The series above is an holomorphic function in the complex plane $z\in \C$  as soon as
the  potential $v$ is stable (see e.g. \cite{Ru}). We remind that a pair potential $v$ is {\it stable} if  there exists  a finite  non-negative number $B_v$ such that
$$
 B_v=\sup_{n\ge 2}~~\sup_{(x_1,\dots,x_n)\in \mathbb{R}^{dn}}\Big\{-{1\over n}\sum_{1\le i<j\le n}v(x_i-x_j)\Big\}\Eq(stabs)
$$
The number  $B_v$ is called the stability constant of the potential $v$.
We will also consider two more  constants associated to the potential $v$. Namely,
$$
\bar B_v=\sup_{n\ge 2}~~\sup_{(x_1,\dots,x_n)\in \mathbb{R}^{dn}}\Big\{-{1\over n-1}\sum_{1\le i<j\le n}v(x_i-x_j)\Big\}\Eq(stabss)
$$
and
$$
B_v^*=\sup_{x\in \Rd}v^-(x)\Eq(bistar)
$$
where $v^-$ is the negative part of the potential, i.e.
$$
v^-(x)={1\over 2}\Big(|v(x)|-v(x)\Big)\Eq(vmen)
$$

\\{\bf Remark}. The constant  $\bar B_v$ is called the {\it Basuev stability constant} of the potential $v$  (after Basuev who was the first to introduce it in \cite{Ba1}) while $B^*_v$ is simply the (absolute value of the) infimum  of the potential.
For a general stable potential $v$ it holds  that
$B_v\le \bar B_v\le {d+1\over d}B_v$   and for any stable potential $v$ in $d\ge 3$ which
reaches a  negative minimum at some  $|x|=r_0$ and is negative for all $|x|>r_0$ it holds 
$\bar B_v\le {2d(d-1)+1\over 2d(d-1)} B_v$ (see \cite{Pr}). For the majority of  potentials used in simulations by  chemists and physicists the stability constant and the Basuev stability constant are likely to be  very close if not equal.
\def\lj{{V_{\,\!\mbox{\tiny\rm LJ}}}}
In particular, for the specific case of the Leonard-Jones potential  $\lj(r)=r^{-12}-2r^{-6}$ in three dimensions
to be considered  later, according to the tables given in \cite{JI}, we have that $B_{\lj}\le \bar B_{\lj}\le {1001\over 1000}B_{\lj}$.
\vskip.2cm
\\Beyond stability, which is the  key property  standing behind  the well-definiteness of the
partition function, a standard hypotheis on the pair potential \cite{Ru}  is the so-called {\it regularity}. Namely,  a pair potential $v$ is   {\it regular} if
$$
\int_{\Rd} |e^{-\b v(x)}-1|dx\doteq C_v(\b) <+\infty\Eq(Cb)
$$
Hereafter we will refer to the constant $C_v(\b)$  as {\it the regularity constant} of the potential.
Note that \equ(Cb)  is equivalent to require that there exists $r_0>0$  such that  $\int_{|x|\ge r_0} |v(x)|dx<+\infty$ (see \cite{Ru}).
All potentials considered  here below are supposed to be stable and regular.

\\In the present paper we will also consider a very relevant subclass of the stable and regular pair potentials which was first proposed  by Basuev in \cite{Ba1}.
This class  is
sufficiently large to embrace the large majority of  examples  of  non purely repulsive pair potential physicists are usually dealing with.
\begin{defi}\label{def9}
A regular  pair potential  $v$  is called
 {\it Basuev} if there exist $\a>0$  such that
$$
v(x)\ge v_\a>0 ~~~~~~~~~~~~~~~~~~~~~~~~~{\rm for ~all}~~~ |x|\le \a\Eq(cb0)
$$
and
$$
v_\a> 2\mu_v(\a)\Eq(cba)
$$
where  $v_\a=\min_{x\in \Rd \atop|x|=\a}v(x)$ and
$$
\mu_v(\a)= \sup_{n\in \mathbb{N},\,(x_1,\dots, x_n)\in \mathbb{R}^{dn}\atop |x_i-x_j|>\a~ \forall \{i,j\}\in E_n}\sum _{i=1}^n
v^-(x_i)
\Eq(mua)
$$
with  $v^-$ being the negative part of the potential defined in \equ(vmen).
\end{defi}
In \cite{Ba1} (see also \cite{dLPY},  Appendix B) it is proved the following useful proposition.
\begin{pro}\label{ba1}
A Basuev potential $v$ according to Definition \ref{def9}  is stable with stability constant $B_v\le {1\over 2} \mu(\a)$.
\end{pro}

\\We stress that many classical potentials utilized in simulations by  physicists and chemists  are Basuev. In particular, as shown in \cite{dLPY}, a   pair potential $v$  of   Lennard-Jones type
(for whose definition and properties we refer the reader to
references \cite{FR}, \cite{Ga}  \cite{Ru}, \cite{MPS}) is Basuev.
%

\vskip.2cm

\\The pressure $P(^vz,\b)$ and the density $\r^v(z,\b)$ of the system under analysis are deduced from the partition function
via the relations
$$
 \b P^v(z,\b)\;=\; \lim_{\L\to \infty}\frac{1}{|\Lambda|}\log\Xi^v_\L(z,\b)\Eq(pressu)
$$
$$
\r^v(z,\b)=   \lim_{\L\to \infty}\frac{ z}{|\Lambda|}{\partial \over \partial z}\log\Xi^v_\L(z,\b)\Eq(densit)
$$
where $|\L|$ denotes the volume of the box $\L$ and here $\lim_{\L\to \infty}$ means that   the size of the cubic box goes to infinity).
It is long known (see \cite{Ru} and references therein) that the limits above exist  whenever $v$ is stable and regular.
It is even longer  known (see e.g. \cite{Mc} and references therein) that, by rewriting $e^{{-\b\sum_{1\le i<j\le n}v(x_i-x_j)}}$ as $\prod_{{1\le i<j\le n}}[e^{ -\b v(x_i-x_j)}-1)+1]$), it is possible to expand the $\log\Xi^v_\L(z)$
as a power series in the fugacity  $z$. Namely,
$$
{1\over |\L|}\log\Xi^v_\L(z)=z+\sum_{n=2}^\infty c^v_n({\Lambda},\b ){z^n}
\Eq(2.3aa)
$$
where, for $n\ge 2$, $c_n({\Lambda},\b )$, the so-called (finite volume) Mayer (a.k.a. Ursell) coefficients  are explicitly given by the formula
$$
c^v_n({\Lambda},\b )~=~{1\over |\L|}{1\over n!}\int_{{\Lambda} }d{x}_1
\dots \int_{{\Lambda} } d{x}_n \sum\limits_{g\in G_n}~
\prod\limits_{\{i,j\}\in E_g}\left[  e^{ -\beta v({x}_i -{x}_j)} -1\right] \Eq(ursm)
$$
with $G_n$  denoting  the set of the connected graphs with vertex set
$[n]\equiv \{1,...,n\}$ and $E_g$ denoting the edge set of $g \in G_n$.
\\Plugging  the expansion \equ(2.3aa) into \equ(pressu) and \equ(densit) we get
$$
 \b P^v(z,\b)\;=\; z+\sum_{n=2}^\infty c^v_n(\b ){z^n}\Eq(pre)
$$
$$
\r^v(z,\b)=  z+\sum_{n=2}^\infty nc^v_n(\b ){z^n}\Eq(dens)
$$
where $c^v_n(\b )= \lim_{\L\to \infty}c^v_n({\Lambda},\b )$ which, since the pair potential $v$ is translational invariant, is given by
$$
c^v_n(\b )=  {1\over n!} \int_{\mathbb{R}^d }d{x}_2
\dots \int_{\mathbb{R}^d } d{x}_n \sum\limits_{g\in G_n}~
\prod\limits_{\{i,j\}\in E_g}\left[  e^{ -\beta v({x}_i -{x}_j)} -1\right] \Eq(ursmi)
$$
Formula \equ(pre) is known as the Mayer series of the pressure.  The coefficients $c^v_n(\b )$ are well defined  (i.e. are all finite) as soon as the potential $v$ is {\it regular}.

\\By inverting the series \equ(dens), i.e. by expressing the fugacity $z$ as power series of the density $\r=\r^v(z,\b)$
(which is always possible, at least in a neighbor of $z=0$, since the first order coefficient of the series \equ(dens) is  non zero),
one can write the pressure of the system in the grand canonical ensemble in power of  $\r$ obtaining the so-called { virial series}, which is usually written as
$$
\b P^v(\r,\b)= \r - \sum_{k\ge 1} {k\over k+1} \b^v_k(\b)\r^{k+1}\Eq(virial)
$$
The equation \equ(virial) is a fundamental one in statistical mechanics: it represents
the equation of state of the (non-ideal) gas of classical particles interacting via the pair potential $v$.
The coefficients $ \b^v_k(\b)$  of the virial series are of course  certain (quite intricate) algebraic combinations  of the Mayer coefficients $c^v_n(\b)$. This algebraic combinations
are known since a long time
(see e.g. formula (49) in \cite{May42} or  also formula (29), p. 319 of \cite{PB} or ) and thus they can in principle be computed just by knowing  the function $v(x)$. The virial  coefficients $\b^v_k(\b)$
also admit
a nice representation in terms of connected graphs  (see e.g.   \cite{Mc} and references therein), namely
$$
\b^v_n(\b )=  \int_{\mathbb{R}^d }d{x}_2
\dots \int_{\mathbb{R}^d } d{x}_n \sum\limits_{g\in G^*_n}~
\prod\limits_{\{i,j\}\in E_g}\left[  e^{ -\beta v({x}_i -{x}_j)} -1\right] \Eq(ursmibi)
$$
where $G^*_n$  is the set of the two-connected graphs with vertex set
$[n]\equiv \{1,...,n\}$. We recall that a graph $g$ is two-connected if deleting any vertex $v$ of $g$ plus the edges incident to $v$ the new graph so obtained remains  connected.

\\A fundamental question in statistical mechanics is to establish the convergence radius $R^v_{\rm Mayer}(\b)$ of the Mayer  series  \equ(pre) and
the convergence radius $\RR^v_{\rm virial}(\b)$ of the virial series \equ(virial).

 \\As far as the Mayer series is concerned,
in 1963
Penrose  obtained \cite{Pe63} the following  upper bound for its coefficients $|c^v_n(\b)|$ when the pair potential $v$ is stable and regular
$$
|c^v_n(\b)| \le ~{n^{n-2}\over n!}e^{2\b B_v(n-2)}[C_v(\b)]^{n-1}\Eq(cnpe)
$$
where $B_v$ is the stability constant of the potetial $v$ defined in \equ(stabs) and $C_v(\b)$ is the regularity constant of the potential $v$
defined in \equ(Cb).
From \equ(cnpe) it  immediately follows that, for a fixed $\b$, of the  series \equ(pre) and \equ(dens) are absolutely convergent
 as soon as $z$ belongs to the open complex disc of radius $R_{\rm PR}(\b)$
centered at the origin  with $R_{\rm PR}(\b)$ given by
$$
R^v_{\rm PR}(\b)=  (1/e) {e^{-2\b B_v}\over C_v(\b)}\Eq(peru)
$$
Therefore we have that  $R^v_{\rm Mayer}(\b)\ge R^v_{\rm PR}(\b)$.
This lower bound   was also  obtained in the same year by
Ruelle \cite{Ru63}.

\\The upper  bound \equ(cnpe) on the absolute value of Mayer coefficients $|c^v_n(\b)|$ for a general stable and regular pair potential   $v$
has been  recently improved by Procacci and Yuhjtman \cite{PY} who proved  that
$$
|c^v_n(\b)| \le ~{n^{n-2}\over n!}e^{\b B_vn}[\tilde C_v(\b)]^{n-1}\Eq(cnpy)
$$
where
$$
\tilde C_v(\b)= \int_{\Rd}( 1- e^{-\b|v(x)|})dx\Eq(cbar)
$$
Bound \equ(cnpy)  forthwith implies that   the Mayer series \equ(pre) of a system of classical particles interacting via a stable and regular  potential
$v$ is absolutely convergent at fixed $\b$ as soon as $|z|< R^v_{\rm PY}(\b)$ where
$$
R^v_{\rm PY}(\b)= {(1/e)}{e^{-\b B_v}\over   \tilde C_v(\b)}\Eq(rpy)
$$
and hence $R^v_{\rm Mayer}(\b)\ge R^v_{\rm PY}(\b)$. Note that, for any $\b>0$, it holds that $R^v_{\rm PY}(\b)\ge R^v_{\rm PR}(\b)$ and the equality  holds only if $B_v=0$. We remind the reader that a stable and
a regular pair potential $v$  such that $B_v=0$ is necessarily non-negative, that is to say,
$v(x)\ge 0$ for all $x\in \Rd$.

\\When the pair potential $v$ is Basuev (i.e. $v$ satisfies Definition \ref{def9}),
 it is possible to obtain an alternative lower bound for the Mayer coefficients using the procedure illustrated by Basuev in \cite{Ba2} and revisited in
 \cite{dLPY}. In these two references the alternative estimate of the Mayer coefficients when Basuev potentials are involoved was given in the following form.
$$
|c^v_n(\b)| \le {n^{n-2}\over n!}\left[{e^{\b\bar{B}_v}-1\over \b\bar{B}_v}\right]^{n-1}\left[\hat{C}_v(\b)
\right]^{n-1}\Eq(A2)
$$
where  $\bar B_v$ is the Basuev stability constant defined in \equ(stabss)  and
$$
\hat{C}_v(\b)= \int_{|x|\le \a} dx ~ \b\,|v(x)|
{\b\bar{B}_v(1-e^{-\b [v(x)-v_\a-\bar{B}_v]})\over \b(v(x)-v_\a-\bar{B}_v)(e^{\b\bar{B}_v}-1)}+ \int_{x\ge \a} dx ~ \b\,|v(x)|\Eq(chat)
$$
with $\a$ and $v_a$  being  the parameters (to be optimized) appearing in the Definition \ref{def9}.
The  estimate \equ(A2) can be rewritten, after simple algebraic manipulations,   in an alternative  and more convenient way for later use. Namely, inequality \equ(A2) can be rewritten as
$$
|c^v_n(\b)| \le {n^{n-2}\over n!}e^{\b \bar B_v(n-1)}[\bm\breve{C}_v(\b)]^{n-1}\Eq(a22)
$$
where
$$
\bm\breve{C}_v(\b)= e^{-\b \bar B_v} \int_{|x|\le \a} dx ~ \b\,|v(x)|
{(1-e^{-\b [v(x)-v_\a-\bar{B}_v]})\over \b(v(x)-v_\a-\bar{B}_v)}+  {1-e^{-\b\bar{B}_v}\over \b\bar B_v}\int_{x\ge \a} dx ~ \b\,|v(x)|\Eq(Cbreve)
$$
Bound \equ(a22) immediately implies that
series \equ(pre) and \equ(dens) are absolutely convergent if $|z|<R_{\rm Ba}(\b)$ where
$$
R^v_{\rm Ba}(\b) =(1/e){e^{-\b\bar B_v}\over \bm\breve{C}_v(\b)}\Eq(radba)
$$
\\Therefore, when $v$ is a Basuev potential according to Definition \ref{def9} we have that $R^v_{\rm Mayer}(\b)\ge R^v_{\rm Ba}(\b)$.
As discussed  in \cite{PY}, in several relevant cases (e.g. if $v$ is the Lennard-Jones potential) the lower bound $R^v_{\rm Ba}(\b)$ for the convergence radius of the Mayer series is larger that the value $R^v_{\rm PY}$ found in \cite{PY}. 

\\To conclude the discussion about the Mayer series, it is  important to stress that the bounds  \equ(cnpy)  and \equ(A2) for  the
 Mayer coefficients are less than or equal to the old Penrose bounds given in the r.h.s. of \equ(cnpe) only if $n\ge 4$. When $n=2,3$ the bound \equ(cnpe) beats both bounds \equ(cnpy) \equ(A2). Of course this has no impact on the
estimate of the  lower bound of the convergence radius of the Mayer series but, as we will see in Section 2 below,  tighter  bounds for $c^v_2(\b)$ and
$c^v_3(\b)$ are crucial  to get efficient estimates of the convergence radius of the virial series.

\\Turning to the virial series \equ(virial),  unfortunately a direct upper bound on its coefficients, as given in \equ(ursmibi), is  so far unavailable. Neverthless, lower bounds for the convergence radius  $\RR^v_{\rm virial}(\b)$ of the virial series \equ(virial)
of a system of particles interacting via a  pair potential $v$ have been obtained during  the last six decades.
The best lower bound for was, until very recently, the one given by Lebowitz and Penrose in 1964 \cite{LP}, who,  assuming that the pair potential $v$ is stable and  regular, proved  that
$$
\RR^v_{\rm virial}(\b)\ge  \RR^v_{\rm LP}(\b)\doteq F(\b B_v){e^{-2\b B_v}\over  C_v(\b)}\Eq(virialold)
$$
where, for $s\ge 0$
$$
F(s)= \max_{w\in (0,1)} {[(1+e^{2s})e^{-w}-1]w\over e^{2s}}\Eq(flp)
$$
The function $F(s)$ is increasing in the interval $s\in [0,+\infty)$
with
$$
F(s=0)=0.144767~~~~~~~~~~~~~~{\rm and} ~~~~~~~~~~~~~~~ \lim_{s\to \infty}F(s)={1\over e}
$$
The bound \equ(virialold) was obtained  via a complex analysis argument using explicitly the  upper bounds on the absolute value of the Mayer coefficients $|c^v_n(\b)|$ given in \equ(cnpe).  Therefore one can reasonably expect  that
the estimates \equ(cnpy) on $|c^v_n(\b)|$, which represent a clear advance with respect to the old  estimates \equ(cnpe),  should  naturally result in an equally strong increase of the lower bound fo $\RR^v_{\rm virial}(\b)$.
Indeed, recently \cite{Pr}  the   Lebowitz-Penrose bound  \equ(virialold) has been  improved for system interacting via a stable and regular pair potential $v$
using  a slight variant
of the bound for $|c^v_n(\b)|$ given in \equ(cnpy). The upper bound for $|c^v_n(\b)|$ proved in \cite{Pr} is as follows.
$$
|c^v_n(\b)| \le ~{n^{n-2}\over n!}e^{\b \bar B_v(n-1)}[\tilde C_v(\b)]^{n-1}\Eq(cnp)
$$
where $\bar B_v$ is the Basuev stability constant defined in \equ(stabss) and $\tilde C_v(\b)$ is the constant defined in \equ(cbar). As shown in \cite{Pr}, the bound \equ(cnp) implies that
$$
\RR^v_{\rm virial}(\b)\ge \RR^v_{\rm Pr}(\b)\doteq  F(0){ e^{-\b \bar B_v}\over   \tilde C_v(\b)}\Eq(prvi)
$$
where $F(0)$ is the function defined in \equ(flp) evaluated at $s=0$.

\\The bound \equ(prvi) represents a very strong improvement on the old Lebowitz-Penrose bound \equ(virialold) when the pair potential $v$ has stability constant $B_v$ strictly positive (i.e $v$   assumes  negative values somewhere), especially for large $\b$.
\\On the other hand, for systems of particles interacting via non-negative potential $v$, for which  $\bar B_v=B_v=0$ and $C_v(\b)=\tilde C_v(\b)$, the bound above coincides with the old  Lebowitz-Penrose bound, namely
$$
\RR^v_{\rm virial}(\b)\ge {0.144767\over C_v(\b)}~~~~~~~\mbox{when $v\ge 0$}\Eq(posiv)
$$
\\In this regard
it is worth to mention that
for the specific case of positive  pair  potentials, the bound \equ(posiv) has very recently been  improved, first by Jansen, Kuna and Tsagkarogiannis \cite{JKT} who replaced the constant
0.144767 by  ${1\over 2e}\approx 0.183939$, and then   by Fern\'andez and Nyoung \cite{NF} who,
inspired on an   unpublished work by S. Ramawadth and S. Tate \cite{RT},  further  improved the same  constant up to the value 0.237961 which
coincides with that claimed (but never  proved) long time ago by Groeneveld \cite{Gr}.

\\Howsoever, these recent results on the convergence radius  of the virial series obtained in \cite{JKT} and \cite{NF},
  as far as non purely repulsive pair potentials are concerned, leave the  situation practically unchanged at the point where it was   after
the paper \cite{Pr}.
The purpose of this note is  to show that, by carefully revisiting   results obtained in papers \cite{dLPY}, \cite{PY}, \cite{Pr} and \cite{Y},  it is possible
to get
new lower bounds of the convergence radius
of the virial series  of systems of classical particles interacting via  stable and tempered pair potentials with non-zero stability constant. These new bounds may become  even better when  Basuev  pair potentials according to Definition \ref{def9} are involved.  To convince the reader of goodness of our new estimates, we pick as an example the  important case of the Lennard-Jones pair potential in $d=3$ and, using  the  accurate estimates obtained in \cite{Y}, we compare
our new bounds with those previously obtained in \cite{Pr}. The main results of this note can be summarized by the following two theorems.

\begin{teorema}\label{pryu}
 Let $v$ be a stable and tempered pair  potential with  Basuev stability constant $\bar B_v$. Then
the convergence radius of the virial  series \equ(virial) of a system of classical particles  at fixed inverse temperature $\b$
interacting via the potential $v$ admits the following lower bound.
$$
\RR^v_{\rm virial}(\b)\ge \RR^v_{\rm stab}(\b)\doteq  F_v^*(\b){ e^{-\b \bar B_v}\over  \tilde C_v(\b)}\Eq(Pvi)
$$
where $\tilde C_v(\b)$ is the constant defined in \equ(cbar),
$$
F_v^*(\b)
=
\max_{w\in (0,1)}w\left[2e^{-w}+(1-e^{-\b (\bar B_v-B_v^*)})we^{-2w}+{3\over 2}
(1-e^{-2\b (\bar B_v-{3\over 2}B_v^*)})w^2e^{-3w}-1\right]\Eq(fstar)
$$
and
$\bar B_v$ and $B_v^*$ are the constants defined in \equ(stabss) and \equ(bistar).
\end{teorema}
Note that $F^*_v(\b)$ is an increasing function in the interval $\b\in [0,+\infty)$. This follows from the fact that, for any stable pair potential $v$ in $d$ dimensions, by definition it holds that $\bar B_v\ge{2}B_v^*$ if $d\ge 3$ and $\bar B_v\ge{3\over 2}B_v^*$ if $d=2$
(and even for the $d=2$ case  it holds  strictly $\bar B_v>{3\over 2}B_v^*$ if the  potential $v$ reaches the  negative minimum $B^*_v$ at some $x_0$ with $|x_0| = r_0$ and is  non-positive for
all $|x| > r_0$). Moreover, by straightforward
calculations (e.g. using Wolframalfa to find the minimum) we have that
$$
F_v^*(\b=0)\approx0.144767~~~~~~~~~\lim_{\b\to +\infty}F_v^*(\b)\approx 0.241857
$$
Note  that in general the function $F^*_v(\b)$ tends to increase rapidly from its minimum value to its maximum value   as $\b$ increases because for most of the
physical potentials the ratio $\bar B_v/B_v^*$ is large  (e.g. is of the order on 10 for the Lennard Jones potentials). Bound \equ(Pvi) given in Theorem \ref{pryu} is  clearly an improvement on  the best lower bound \equ(prvi) obtained in \cite{Pr} for convergence radius of the virial series as far as stable and regular potentials with strictly positive stability constant are concerned . On the other hand,  when  positive potentials are considered, for which  $\bar B_v=B_v=0$ and $C_v(\b)=\tilde C_v(\b)$, the bound \equ(Pvi) above, similarly to bound \equ(prvi), coincides with Lebowitz-Penrose bound \equ(posiv).

\begin{teorema}\label{basu}
Let $v$ be a Basuev potential according to Definition \ref{def9}.
Then
the convergence radius of the virial  series \equ(virial)  of a system of classical particles interacting via the potential $v$ admits the following lower bound.
$$
\RR^v_{\rm virial}(\b)\ge \max\Big\{\RR^v_{\rm stab}(\b),\RR^v_{\rm Bas}(\b)\Big\}\Eq(PYvi0)
$$
where  $ \RR^v_{\rm stab}(\b)$ is defined in \equ(Pvi) and 
$$
\RR^v_{\rm Bas}(\b)\doteq  \bm\breve F_v(\b){ e^{-\b \bar B_v}\over \bm\breve C_v(\b) }\Eq(PYvi)
$$
with
$\bm\breve{C}_v(\b)$ being  defined in \equ(Cbreve) and
$$
\bm\breve F_v(\b)=   \max_{w\in (0,1)}
 w\Bigg[2e^{-w}+\left(1-e^{-\b (\bar B_v-B_v^*)}{\tilde C_v(\b)\over \bm\breve C_v(\b)}\right)we^{-2w}+
 $$
 $$
~~~~~~~~~~~~~~~~~~~~+{3\over 2}
\left(1-e^{-\b (2\bar B_v-{3}B_v^*)}\left[{ \tilde C(\b)\over \bm\breve  C_v(\b)}\right]^2\right)w^2e^{-3w}-1\Bigg]\Eq(hatf)
$$
\end{teorema}
Note that the quantity $\bm\breve F_v(\b)$ is always strictly greater than zero since for any real $c_1$ and $c_2$ the function
$f(w)= w([2e^{-w}+c_1 we^{-2w}+c_2 w^2e^{-3w}-1)$ has always a positive maximum in the interval $w\in (0,1)$.
We will see ahead  through a popular example (the Lennard-Jones fluid)  that  for systems of particles interacting via a Basuev potential
we may have that $\RR_{\rm Ba}^v(\b)>\RR^v_{\rm stab}(\b)$.


\vskip.51cm

\numsec=2\numfor=1

\subsection*{2. Proof of Theorem 1}\label{subsec3}

First of all we want to point out that in doing  the estimate of the convergence radius of the virial series
 {\it a la} Lebowitz-Penrose as described below, it is crucial to have
upper bounds for $|c_n(\beta)|$ defined in \equ(ursmi)  as tight as possible when $n=2$ and $=3$. We thus prove the following preliminary lemma.
\begin{lem}\label{lem} Given  a regular pair potential $v$, let $c^v_n(\b)$ be  defined as in  \equ(ursmi). Then the following bounds hold.
$$
|c^v_2(\beta)|
\le {1\over 2}e^{\b B^*_v}\tilde C_v(\beta)\Eq(Proyuh3)
$$
$$
|c^v_3(\beta)|
\le {1\over 2}
\left[e^{ {3\over2}\b B_v^*}\tilde C_v(\beta)\right]^2\Eq(Proyuh3b)
$$
\end{lem}
{\bf Proof}. Let first analyze the case $n=2$.  By \equ(ursmi) we have that
$$
|c^v_2(\b)|= {1\over 2!}\left|\int_{\Rd}(e^{-\b v(x)}-1)dx\right|\le  {1\over 2!}\int_{\Rd}\left|e^{-\b v(x)}-1\right|dx \Eq(ass)
$$
Now observe that
$$
\left|e^{-\b v(x)}-1\right|= \cases{ (1-e^{-\b |v(x)|}) &if $v(x)\ge 0$\cr\cr
e^{-\b v(x)}(1-e^{-\b |v(x)|}) & if $v(x)< 0$
}
$$
and since $e^{-\b v(x)}\le e^{+\b B^*_v}$ for all $x\in \Rd$ by definition of $B^*_v$, we have in any case
$$
\left|e^{-\b v(x)}-1\right|\le e^{+\b B^*_v}(1-e^{-\b |v(x)|}) \Eq(iss)
$$
Plugging \equ(iss) into \equ(ass) we get \equ(Proyuh3).

\\Let us now consider tha case $n=3$. By \equ(ursmi) we have
$$
|c^v_3(\b)|={1\over 3!}\Bigg|3\int_{\Rd}\!\!\!\!dx\!\!\int_{\Rd}\!\!\!\!dy(e^{-\b v(x)}-1)(e^{-\b v(y)}-1)
+\int_{\Rd}\!\!\!\!dx\!\!\int_{\Rd}\!\!\!\!dy(e^{-\b v(x)}-1)(e^{-\b v(y)}-1)(e^{-\b v(x-y)}-1)
\Bigg|=
$$
$$
= {1\over 3!}\Bigg|2\int_{\Rd}\!\!\!\!dx\int_{\Rd}\!\!\!\!dy(e^{-\b v(x)}-1)(e^{-\b v(y)}-1)+
\int_{\Rd}\!\!\!\!dx\int_{\Rd}\!\!\!\!dy(e^{-\b v(x)}-1)(e^{-\b v(y)}-1)e^{-\b v(x-y)}\Bigg|\le
$$
$$
\le {1\over 3!}\Bigg(2\int_{\Rd}\!\!\!\!dx\int_{\Rd}\!\!\!\!dy|e^{-\b v(x)}-1||(e^{-\b v(y)}-1|+
\int_{\Rd}\!\!\!\!dx\int_{\Rd}\!\!\!\!dy|e^{-\b v(x)}-1||e^{-\b v(y)}-1|e^{-\b v(x-y)}\Bigg)
$$
Therefore, using again \equ(iss) we get
$$
|c^v_3(\b)|\le {1\over 3!}\left[ 2e^{2\b B^*_v}[\tilde C_v(\b)]^2+e^{3\b B^*_v}[\tilde C_v(\b)]^2\right]\le
{1\over 2}e^{3\b B^*_v}[\tilde C_v(\b)]^2= {1\over 2}[e^{{3\over 2}\b B^*_v}\tilde C_v(\b)]^2
$$
and the proof of Lemma \ref{lem} is concluded. $\Box$
\vskip.2cm

\\For $n\ge 3$, we will   bound $|c^v_n(\beta)|$ either as in \equ(cnp) if $v$ is stable and regular or as in  \equ(a22) if $v$  is Basuev.

\\We now follow the steps done by Lebowitz and Penrose in \cite{LP} to obtain a lower bound for $\RR^v_{\rm virial}(\b)$. We start by observing that, due to \equ(dens) and the fact that $c_1(\b)=1$, there exists a circle $C$
 of some radius $R< 1/[\tilde C(\b)e^{\b B_v+1}]$ and center in the origin  $z=0$ of the complex $z$-plane such that
  $\r^v(\b,z)$
has only one zero  in the disc ${ D_R}=\{z\in \C: |\l|\le R\}$ and this zero occurs precisely at $z=0$.
  Let now  $\r\in \C$ be such that
$$
|\r|< \min_{z\in C}|\r^v(\b,z)| \Eq(condm)
$$
Then by Rouch\'e's Theorem $\r^v(\b,z)$ and
 $\r^v(\b,z)-\r$ have the same number of zeros (i.e. one)  in the region $D_R=\{z\in \C: |z|\le R\}$. In other words, for any complex $\r$ satisfying
  \equ(condm) there is only one $z\in D_R$ such that  $\r=\r^v(\b,z)$ and therefore we can invert the equation   $\r=\r^v(\b,z)$ and write
 $z=z(\b, \r)$. Thus, according to  Cauchy's argument principle, we can write the pressure $\b P^v(\b,z)$
  as a function of the density $\r=\r(\b,z)$  as
$$
P^v(\r,\b)={1\over 2\p i}\oint_\g P^v(\b,z){d\r^v(\b,z)\over dz}{dz\over \r^v(\b,z)-\r}\Eq(cauchy)
$$
where $\g$ can be  any circle centered at the origin in the  complex plane fully contained in the region $D_R$
and such that
$$
|\r|<\min_{z\in \g} | \r^v(\b,z)|\Eq(minro)
$$
The function  $P(\r,\b)$ is clearly analytic in $\r$ in the region \equ(minro). Indeed,
once \equ(minro) is satisfied we can write
$$
{1\over \r^v(\b,z)-\r}=\sum_{n=0}^{\infty} {\r^n\over  [\r^v(\b,z)]^{n+1}}\Eq(okso)
$$
and inserting \equ(okso) in \equ(cauchy) we get
$$
P^v(\r,\b)= \sum_{n=1}^\infty k_n(\b)\r^n\Eq(virial3)
$$
with
$$
k_n(\b)= {1\over 2\p i}\oint_\g P^v(\b,z){d\r^v(\b,z)\over dz}{dz\over  [\r^v(\b,z)]^{n+1}} ~=
~
~-{1\over 2\p in}\oint_\g P^v(\b,z){d\over dz}\left[{1\over  [\r^v(\b,z)]^{n}}\right]dz~=
$$
$$
= {1\over 2\p in}\oint_\g {dP^v(\b,z)\over dz}{1\over  [\r^v(\b,z)]^{n}}dz~= ~ {1\over 2\p in\b} \oint_\g{1\over
[\r^v(\b,z)]^{n-1}}{dz\over z}~~~~~~~~~~~~~~~~~~~~~
$$
Therefore
$$
|k_n(\b,\L)|\le{1\over n\b}{1\over \left[\min_{z\in \g} |\r^v(\b,z)|\right]^{n-1}}\Eq(cn)
$$
Inequality \equ(cn) shows that the convergence radius $\RR^v_{\rm virial}(\b)$ of the series \equ(virial3) (i.e. of the virial series \equ(virial)) is such
that
$$
\RR^v_{\rm virial}(\b)\ge  \max_{r\in D_R} \min_{z\in \C\atop|z|=r}   |\r^v(\b,z)|\Eq(Ok)
$$
The game is thus to find  a  circle $\g$ of optimal radius $r_\g$ in the region $D_R$ in such a way to maximize the r.h.s. of \equ(Ok).
We proceed as follows.
Recalling \equ(dens)  we have, by the triangular inequality, that
$$
 |\r^v(\b,z)| \ge |z|- \sum_{n=2}^\infty n|c^v_n(\b)||z|^n \Eq(triangul)
$$
We now  use estimates \equ(Proyuh3) and \equ(Proyuh3b) and  to bound the first two coefficients of the sum in the r.h.s. of  \equ(triangul) and estimates \equ(cnp) for the remaining coefficients.
Therefore we get
$$
  \max_{r\in D_R} \min_{z\in \C\atop|z|=r} \!|\r^v(\b,z)| \ge \max_{r\in D_R}\Big\{r-  e^{\b B_v^*}\tilde C_v(\beta)r^2-{3\over 2}
\left[e^{ {3\over2}\b B_v^*}\tilde C_v(\beta)\right]^2r^3-\sum_{n=4}^\infty {n^{n-1}\over n!}[\tilde C_v(\b) e^{\b \bar B_v}]^{n-1}r^n\Big\}=
 $$
 $$
 =
  \max_{r\in D_R}\!\!\Big\{2r+\tilde C_v(\b)( e^{\b \bar B_v}\!- e^{\b B_v^*})r^2+{3\over 2}[\tilde C_v(\b)]^2(
 e^{2\b \bar B_v}\!-e^{3\b B_v^*})r^3-{1\over \tilde C_v(\b) e^{\b \bar B}}\!\!\sum_{n=1}^\infty {n^{n-1}\over n!}[\tilde C_v(\b) e^{\b \bar B_v}r]^{n}\!\!\Big\}\!\ge
$$
$$
 \ge  \max_{x\in (0,{1\over e})} {1\over \tilde C_v(\b) e^{\b \bar B_v}}\left[2x+(1-e^{-\b (\bar B_v-B_v^*)})x^2+{3\over 2}
(1-e^{-2\b (\bar B_v-{3\over 2}B_v^*)})x^3-\sum_{n=1}^\infty {n^{n-1}\over n!}x^{n}\right]\Eq(rhs)
$$
where we have set  $x= e^{\b \bar B_v} \tilde C_v(\b)r$  and we have taken  $x\in (0,{1/e})$ which
surely  inside the convergence region since $\bar B_v\ge B_v$ and $\tilde C_v(\b) e^{\b \bar B_v}|\l|<1/e$.
The r.h.s. of \equ(rhs) can be written in a closed form by the change of variables
$$
x=we^{-w}
$$
which is one-to one as  $x$ varies in the interval $[0, {1\over e}]$ and  $w$ varies in the interval $[0,1]$. Indeed using Euler formula
$$
w=\sum_{n=1}^\infty{n^{n-1}\over n!}(we^{-w})^n\Eq(eule)
$$
valid for all $w\in [0,1]$.  We thus
can rewrite  the r.h.s. of \equ(rhs) as
$$
 {1\over \tilde C_v(\b) e^{\b \bar B_v} } \max_{w\in (0,1)}\Bigg\{ w\,\Big[2e^{-w}+(1-e^{-\b (\bar B_v-B_v^*)})we^{-2w}+{3\over 2}
(1-e^{-2\b (\bar B_v-{3\over 2}B_v^*)})w^2e^{-3w}-1\Big]\Bigg\}
$$

\\In conclusion we get that
$$
\RR^v_{\rm virial}(\b)\ge {F^*_v(\b)\over \tilde C_v(\b) e^{\b \bar B_v}}\Eq(Ok2)
$$
where
$F^*_v(\b)$ is precisely the constant defined in \equ(fstar).

\subsection*{3. Proof of Theorem 2}
\numsec=3\numfor=1
The proof of Theorem \ref{basu} can be done along the same lines of the previous section. 
First of all, since a Basuev potential $v$  is  stable and regular, by Theorem \ref{pryu} we have that 
$\RR^v_{\rm virial}(\b)\ge \RR^v_{\rm stab}(\b)$. Thus to prove \equ(PYvi0)  we need to show that  it also holds that
$\RR^v_{\rm virial}(\b)\ge \RR^v_{\rm Ba}(\b)$.
We use once again
estimates \equ(Proyuh3) and \equ(Proyuh3b)  to bound the first two coefficients of the sum in the r.h.s. of  \equ(triangul) but  now we use estimates \equ(a22) for the remaining coefficients.
We get
$$
 |\r^v(\b,z)| ~\ge ~|z|-  e^{\b B_v^*}\tilde C_v(\beta)|z|^2-{3\over 2}
\left[e^{ {3\over2}\b B_v^*}\tilde C_v(\beta)\right]^2|z|^3-\sum_{n=4}^\infty {n^{n-1}\over n!}
\left[\bm\breve C_v(\b) e^{\b\bar{B}_v}\right]^{n-1}|z|^n~=~
 $$
 $$
 =~
 2|z|+\left[\bm\breve C_v(\b) e^{\b\bar{B}_v}- e^{\b B_v^*}\tilde C_v(\beta)]\right]|z|^2+
 {3\over 2}\left[\left[\bm\breve C_v(\b) e^{\b\bar{B}_v}\right]^2-\left[e^{3\b B_v^*\over 2}\tilde C_v(\beta)\right]^2\right]|z|^3-
 $$
 $$
 -{1\over \bm\breve C_v(\b) e^{\b \bar B_v}}\sum_{n=1}^\infty {n^{n-1}\over n!}\left[\bm\breve C_v(\b) e^{\b\bar{B}_v}|z|\right]^{n}
$$
i.e. we can bound
$$
 \max_{r\in D_R} \min_{z\in \C\atop|z|=r}|\r^v(\b,z)| \ge  {1\over \bm\breve C_v(\b) e^{\b \bar B_v}}
  \max_{x\in (0,{1\over e})}\Bigg\{
 \Bigg[2x+\left((1-e^{-\b (\bar B_v-B_v^*)}{\tilde C_v(\b)\over\bm\breve  C_v(\b)}\right)x^2+
 $$
 $$
 +{3\over 2}
\left(1-e^{-\b (2\bar B_v-{3}B_v^*)}\left[{ \tilde C_v(\b)\over\bm\breve C_v(\b)}\right]^2\right)x^3-\sum_{n=1}^\infty {n^{n-1}\over n!}x^{n}\Bigg]\Bigg\}\Eq(rhs3)
$$
where we have set  $x= \bm\breve C(\b) e^{\b\bar{B}_v}r$  and   $x\in (0,{1/e})$ is
  inside the convergence region due to \equ(radba).
Setting as before
$x=we^{-w}$
and using again  Euler formula \equ(eule)
 we
get
$$
 \max_{r\in D_R} \min_{z\in \C\atop|z|=r}|\r^v(\b,z)| \ge  {1\over \bm\breve C_v(\b) e^{\b \bar B_v}} \max_{w\in (0,1)}\Bigg\{
 w\Bigg[2e^{-w}+\left((1-e^{-\b (\bar B_v-B_v^*)}{\tilde C(\b)\over\bm\breve  C_v(\b)}\right)we^{-2w}+
 $$
 $$
 +{3\over 2}
\left(1-e^{-\b (2\bar B_v-{3}B_v^*)}\left[{ \tilde C_v(\b)\over\bm\breve  C_v(\b)}\right]^2\right)w^2e^{-3w}-1\Bigg]\Bigg\}
$$

\\In conclusion we get that
$$
\RR^v_{\rm virial}(\b)\ge {1\over \bm\breve C_v(\b) e^{\b \bar B_b}}
\bm\breve F_v(\b)\Eq(Ok3)
$$
with $\bm\breve F_v(\b)$ coinciding with the function defined in \equ(hatf) and therefore \equ(PYvi0) is proved.

\subsection*{4. An application: the three-dimensional Lennard-Jones potential}\label{sec4}
\numsec=4\numfor=1
Let us compare in this final section our new bounds $\RR^v_{\rm stab}(\b)$ defined in \equ(Pvi) and $\RR^v_{\rm Bas}(\b)$ defined in \equ(PYvi)
 with the
bound  $\RR^v_{\rm PY}(\b)$  given in  \equ(prvi) for the convergence radius of the virial series  of  systems of particles interacting through a pair potential $v$ with strictly positive
stability constant.

\\We will  specifically examine the case of the Lennard-Jones potential,
usually written as $v(r)= 4\e[\left({\s/r}\right)^{12}-  \left({\s/ r}\right)^{6}]$
with $\e$ being  the depth of the potential well and  $\s$ being  the distance at which the potential is zero.  This potential  is by far the most used to model interaction between molecules in simulations by chemists and physicists.
As showed in \cite{dLPY}, the Lennard-Jones potential
is Basuev (see Proposition 2 in \cite{dLPY}), so we are free to
use also the  estimate \equ(PYvi0)   for $\RR^v_{\rm virial}(\b)$.

\\By suitably rescaling inverse temperature and distances,  we may assume without loss of generality that the   Lennard-Jones potential has the following expression
$$
\lj(x)= {1\over |x|^{12}}-  {2\over |x|^6}\Eq(sclj)
$$
We remind the reader that we considering here only the three-dimensional case, so in \equ(sclj) it is understood that  $x\in \mathbb{R}^3$. Moreover, for simplicity, we will  do the calculations
setting the (rescaled) inverse temperature at the value  $\b=1$.

\\First of all it is known that the stability constant $B_{\lj}$  of  the three-dimensional  rescaled Lennard-Jones potential $V_{\rm LJ}$ given in \equ(sclj) is bounded as follows.
$$
8.61\le B_{\lj}\le 14.316
$$
The current  lower bound has been obtained in \cite{SABS} while the current upper bound has been obtained in \cite{Y}.
Concerning the  estimate of the Basuev stability constant $\bar{B}_{\lj}$ of the potential $\lj$,  we
use the data made avaliable in \cite{cambri} which show
that  $ \bar{B}_{\lj}\le {(1.001)} B_{\lj}$ and thus,
$$
8.61\le \bar B_{\lj}\le 14.331\Eq(serglj)
$$
In regard to value of  $\tilde C_{V_{\rm LJ}}(\b=1)$ defined in \equ(cbar) to be plugged in
\equ(prvi),
a straightforward (computer assisted) calculation gives
$$
\tilde C_{V_{\rm LJ}}(\b=1) =4\pi\int_0^{\infty}r^2(1-e^{-{|{1\over r^{12}}-{2\over r^6}|}})dr\approx 9.1864 \Eq(cblj)
$$
To compute $\RR^{\lj}_{\rm Bas}(\b=1)$  we will need
to estimate $\bm\breve{C}_{\lj}(\b)$ defined in \equ(Cbreve) at $\b=1$.
Observe that
for any Basuev potential $v$
and
any $\a$ satisfying conditions \equ(cb0) and \equ(cba), the quantity 
$$
e^{-\b\bar{B}_v}{(1-e^{-\b [v(x)-v_\a-\bar{B}_v]})\over \b(v(x)-v_\a-\bar{B}_v)}
$$
appearing in the integrand of the first term of the r.h.s. of \equ(Cbreve) is monotone decreasing as a function of $\bar B_v$ for any $v(x)-v_a\ge 0$. So, recalling that 8.61 is a lower bound for $\bar B_\lj$,  we have  that
$$
e^{-\bar{B}_\lj}{(1-e^{-[\lj(x)-\lj_\a-\bar{B}_\lj]})\over (\lj(x)-v_\a-\bar{B}_\lj)}\le 
e^{-8.61}{(1-e^{-[\lj(x)-\lj_\a-8.61]})\over (\lj(x)-\lj_\a-8.61)}
$$
where
$$
\lj_a={1\over \a^{12}}-{2\over \a^6}
$$
On the other hand the quantity $(1-e^{-\b\bar B_v})/\bar B_v$ multiplying the second integral in the r.h.s. of \equ(Cbreve) is also
decreasing as a  function of $\bar B_v$.
Therefore, once the constant $\a>0$ satisfying conditions
\equ(cb0) and \equ(cba) in Definition \ref{def9} has been established, we can bound $\hat{C}_{V_{\rm LJ}}(\b=1,  \bar{B}_{\lj})$ as follows
$$
\bm\breve{C}_{V_{\rm LJ}}(\b=1) \le
e^{-8.61}\int_{\mathbb{R}^3:\,|x|\le \a}
\!\!\!\!dx \,\lj(|x|)
{(1-e^{-[\lj(x)-\lj_\a-8.61]})\over (\lj(x)-\lj_\a-8.61)}+ {(1-e^{-8.61})\over 8.61}\int_{\mathbb{R}^3:\,|x|\ge \a} \!\!\!\!dx\,|\lj(x)| \Eq(hatClj)
$$
The game now is to find  a clever choice of  the parameter $\a$  for the Lennard-Jones potential $\lj$. That is to say, $\a$ should be a number which satisfies    \equ(cb0) and \equ(cba)  and  at the same time makes the r.h.s. of \equ(hatClj) as small as possible. This can be achieved
by using a recent result obtained by Yuhjtman specifically of the Lennard-Jones potential \cite{Y}.
Namely, Yuhjtman proved (see Proposition 3.1 in \cite{Y})  that for Lennard-Jones potential  $\lj$ given by \equ(sclj),  taking  $\a\in [0.6, 0.7]$,  the number
$\m_{_{\lj}}(\a)$  defined in \equ(mua) is such that
$$
\m_{_{\lj}}(\a)\le  {24.05\over a^3}
$$
This  immediately implies that $\lj(\a)\ge 2\mu(\a)$ as soon as $\a\le 0.6397$. Thus, by choosing $\a\in  [0.6, 0.6397]$, the Lennard-Jones potential \equ(sclj) satisfies conditions \equ(cb0) and \equ(cba) in Definition \ref{def9}.
Taking  $\a=0.6397$, which is the value in the interval $[0.6, 0.6397]$ that minimizes the r.h.v. of \equ(hatClj),  we get
$$
\bm\breve{C}_{\lj}(\b=1)~\le~ {4\pi e^{- 8.61}}\int_0^{0.6397} dr~ \left[{1\over r^{10}}-{2\over r^4}\right]
{(1-e^{- \left[{1\over r^{12}}-{2\over r^6}-{1\over |0.6397|^{12}}+{2\over |0.6397|^6}-8.61\right]})\over
({1\over r^{12}}-{2\over r^6}-{1\over |0.6397|^{12}}+{2\over |0.6397|^6}-8.61)}~~~+
$$
$$
~~~~~~~~~~~~~~~~+~4{(1-e^{-8.61})\over 8.61}\pi\int_{0.6397}^\infty dx~ \left|{1\over r^{10}}-{2\over r^4}\right|~\le~~0.0345876+ 7.15664~~\le~~7.2\Eq(chatlj)
$$
on the other hand, recalling \equ(serglj), we have also that
$$
\bm\breve{C}_{\lj}(\b=1)~\ge~ 4{(1-e^{-14.331})\over 14.331}\pi\int_{0.6397}^\infty dx~ \left|{1\over r^{10}}-{2\over r^4}\right|\ge 4.3
$$
We also need to evaluate  the quantities $F^*_{V_{\rm LJ}}(\b)$ and $\hat F_{V_{\rm LJ}}(\b)$ defined in \equ(fstar) and \equ(hatf) when $\b=1$.
Considering that $B_{V_{\rm LJ}}^*=1$  and that $\bar B_{\rm LJ}\ge 8.61$ for the Lennard-Jones potential \equ(sclj), we can  estimate
$$
F_{V_{\rm LJ}}^*(\b=1)\ge
\max_{w\in (0,1)}w\left[2e^{-w}+(1-e^{-(7.61)})we^{-2w}+{3\over 2}
(1-e^{-(14.22)})w^2e^{-3w}-1\right]
\ge 0.2418
$$
On the other hand, the quantity $\bm\breve F_\lj(\b=1)$ can be bounded below, as far as the rescaled Lennard-Jones potential is concerned,by the following value.
$$
\bm\breve F_{V_{\rm LJ}}(\b=1)\ge  \max_{w\in (0,1)}\Bigg\{
 w\Bigg[2e^{-w}+\left(1-e^{-(7.61)} {9.2\over 4.3}\right)we^{-2w}+
 $$
 $$
 ~~~~~~~~~~~~~~~~~~~~~~~~~~~~~~+
 {3\over 2}
\left(1-e^{-(14.22)}\left({9.2\over4.3}\right)^2\right)w^2e^{-3w}-1\Bigg]\Bigg\}\Eq(hatf2)
$$
A straightforward (computer assisted) computation gives 
$$
\bm\breve F_{V_{\rm LJ}}(\b=1)\ge 0.2417
$$

\\We can now compare the best previous lower bound  $\RR^\lj_{\rm Pr}(\b)$ given in  \equ(prvi)
for the  convergence radius $\RR^\lj_{\rm virial}(\b)$ of the virial series  of a system of particles at $\b=1$ interacting via the rescaled Lennard-Jones potential given in \equ(sclj) with the new bounds
$\RR^\lj_{\rm stab}(\b)$ and  $\RR^\lj_{\rm Ba}(\b)$  produced by \equ(Pvi) and \equ(PYvi). First, according to bound \equ(prvi) given in \cite{Pr} we have
$$
\RR^\lj_{\rm Pr}(\b=1)\approx   {0.144767\over 9.1864} e^{-\bar B_{\rm LJ}}\approx 0.015759e^{-\bar B_{\rm LJ}} \approx {e^{-\bar B_{\rm LJ}}\over 63.45}\Eq(Pvil)
$$
On the other hand, the new bound   \equ(Pvi) given in Theorem \ref{pryu} yields
$$
\RR^\lj_{\rm stab}(\b=1)\approx   {0.2418\over 9.1864} e^{-\bar B_{\rm LJ}}=0.02632e^{-\bar B_{\rm LJ}} \approx {e^{-\bar B_{\rm LJ}}\over 38}\Eq(Pvil2)
$$
which represent an improvement by a factor 1.67 with respect to bound \equ(Pvi).

\\An ever better bound is obtained using the bound \equ(PYvi) given in Theorem \ref{basu}. Doing so we get
$$
\RR^\lj_{\rm Ba}(\b=1)
\ge   {0.2417\over 7.2}e^{-\bar B_{\rm LJ}}\ge 0,03369 e^{-\bar B_{\rm LJ}}
\ge {e^{-\bar B_{\rm LJ}}\over 29.8}\Eq(PYvi2)
$$
which represents an improvement more than twice  better than  bound  \equ(Pvil).

\subsection*{Acknowledgements} This work has been partially supported by the Brazilian agencies
Coordenadoria de Aperfei\c{c}oamen-to de Pessoal de N\'\i vel Superior (CAPES) and
 Conselho Nacional de Desenvolvimento Cient\'\i fico e Tecnol\'ogico (CNPq).

\vskip.55cm
\renewcommand{\theequation}{A.\arabic{equation}}
\setcounter{equation}{0}  

\vskip.55cm

\renewcommand{\section}[2]{}%

\end{document}